\documentclass[aps,prd,showpacs,floatfix,twocolumn,nofootinbib,superscriptaddress]{revtex4-2} %
\usepackage{dcolumn}

\usepackage{amsmath,amsfonts,amssymb,mathrsfs,times,bm}
\usepackage{extarrows}
\usepackage{float}
\usepackage{booktabs}
\usepackage{graphicx,epstopdf}
\usepackage{dcolumn}
\usepackage{exscale}
\usepackage{relsize}
\usepackage{latexsym}
\usepackage{longtable}
\usepackage[colorlinks=true,breaklinks=true,linkcolor=blue,citecolor=blue,urlcolor=blue]{hyperref}
\usepackage{url}
\usepackage[]{units}
\newcommand{\nn}{\nonumber}

\newcommand{\orcid}[1]{\href{https://orcid.org/#1}{\includegraphics[scale=0.035]{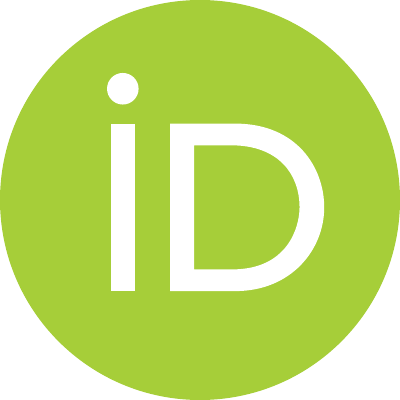}}}

\begin{document}

\title{Gravitational deflection of light in polar-axis plane of a moving Kerr-Newman black hole}

\author{Xuan Wang}
\affiliation{School of Mathematics and Physics, University of South China, Hengyang 421001, China}
\author{Wenbin Lin\hspace*{0.6pt}\orcid{0000-0002-4282-066X}\hspace*{0.8pt}}
\affiliation{School of Mathematics and Physics, University of South China, Hengyang 421001, China}
\affiliation{School of Physical Science and Technology, Southwest Jiaotong University, Chengdu 610031, China}
\author{Bo Yang}
\affiliation{School of Mathematics and Physics, University of South China, Hengyang 421001, China}
\affiliation{Purple Mountain Observatory, Chinese Academy of Sciences, Nanjing 210023, China}
\author{Guansheng He\hspace*{0.6pt}\orcid{0000-0002-6145-0449}\hspace*{0.8pt}}
\email{Corresponding author. E-mail: hgs@usc.edu.cn}
\affiliation{School of Mathematics and Physics, University of South China, Hengyang 421001, China}
\affiliation{Purple Mountain Observatory, Chinese Academy of Sciences, Nanjing 210023, China}

\date{\today}

\begin{abstract}
The gravitational deflection of light signals restricted in the polar-axis plane of a moving Kerr-Newman (KN) black hole with a constant velocity along the polar axis is studied within the second post-Minkowskian (PM) approximation. For this purpose, the Lorentz boosting technique is adopted to obtain the exact metric of a moving KN black hole with an arbitrary constant velocity in Kerr-Schild coordinates for the first time. Based on the weak field limit of the exact metric, we then derive the equations of motion of test particles constrained in the polar-axis plane of a moving KN source whose velocity is along the polar axis and collinear with its angular momentum. An iterative technique is utilized subsequently in the calculations of the null deflection angle up to the 2PM order caused by the moving lens, and this deflection angle is found to be spin-independent. Finally, we discuss the influence of the motion of the lens on the gravitational deflection and estimate the possibility to detect this kinematical effect. Our work might be helpful for future astronomical observations.

\begin{description}
\item[Keywords]
Black holes; Exact metric; Gravitational lens
\end{description}

\end{abstract}

\maketitle

\section{Introduction}  \label{sect1}
Gravitational lensing, along with its extensive applications to many fields in astronomy, has developed into one of the most rapidly growing branches of modern astrophysics (see, for instance,~\cite{Refsdal1964,BN1992,Wambsganss1998,VE2000,Bozza2002,KP2005,Virbha2009,Reyes2010,Liu2016,ZX2016,LYJ2016,FLBPZ2017,ZX2017E,Collett2018,JOSVG2018,
JS2019,LX2019E,MLMHBN2019,LMWX2019,ZX2020a,CLCS2020,MWS2020b,LX2021,GX2021,BAB2021,CX2021,ZX2022,GX2022,LLQZ2022,JRPO2022,Virbhadra2022,SPVR2023,SVP2023,VBG2024,PAJJA2024,HYAJA2024,Virbhadra2024}). Different from much literature with the focus on the gravitational lensing phenomena in a static spherically symmetric or stationary axially symmetric spacetime,
the gravitational lensing effects caused by a moving one-body gravitational system or by a N-body system consisting of moving gravitational sources have been the subject of quite a few previous works in the last decades (see, e.g.,~\cite{BG1983,KS1999,KP2003,WS2004,Heyrovsk2005,KM2007,KF2007,DX2012,HL2014,BZ2015,Deng2015,HL2016a,HL2017b,Zscho2018,Zscho2019,LFZMH2021} and references therein). For example, a pioneering method on the basis of the influence of the lens' motion on the brightness of an isotropic radiation field was proposed to measure the transversal velocity of a cluster of galaxies~\cite{BG1983}. Wucknitz and Sperhake~\cite{WS2004} investigated the effects of the radial and transversal motions of a Schwarzschild lens on the gravitational deflection of test particles including photons within the framework of the first post-Minkowskian (PM) approximation. The equatorial bending phenomena of relativistic massive particles and light in the spacetime of a radially moving Kerr-Newman (KN) black hole were also studied~\cite{HL2016a,HL2017b}, with a discussion on the effects of the len's motion on the second-order contributions to the deflection angle. The reasons responsible for an increasing interest on this type of gravitational lensing effects lie mainly in two aspects. On the theoretical side, the so-called velocity effect~\cite{PB1993,WS2004,Heyrovsk2005} resulted from the motion of the gravitational source has an influence on propagation of test particles, and may eventually affect the related observable relativistic effects. It is interesting and necessary to further probe the velocity effects on the lensing properties for various scenarios in different spacetimes. On the observational aspect, it should be mentioned that those velocity effects for the cases of the lens motions with a relatively large or relativistic velocity may be so evident that they could play a role in the measurements of a part of lensing observables~\cite{WS2004,Heyrovsk2005,HL2016a}. As is known, high-velocity or hypervelocity celestial bodies (such as stars, neutron stars, or even stellar-mass black holes) are not rare in our universe. For example, the transverse velocity of the pulsar B$2224+65$ is not smaller than $800\,$km/s~\cite{CRL1993}, while the pulsar B$1508+55$ moves with a high transverse velocity $\sim 1083^{+103}_{-90}\,$km/s~\cite{CVBCCTFLK2005}. When such a high-velocity celestial body serves as the gravitational lens, a full consideration of the lens motion effects on both the first-order contributions and the higher-order contributions to the lensing properties (e.g., the differential time delay of the primary and secondary images of the source) is necessary in future high-accuracy astronomical observations, as discussed in~\cite{HL2016a}. Additionally, it is known that rapid improvement of the high-accuracy astronomical instruments and techniques has been achieved in the past decades. Current and forthcoming surveys with multiwavelength observations are working towards an astrometric precision of $1\!\sim\!10$ microarcseconds ($\mu$as) or better~\cite{Perryman2001,Prusti2016,SN2009,Reid2009,Chen2014,Malbet2012,RH2014,Malbet2014,Trippe2010,ZRMZBDX2013,Murphy2018,RD2020,Brown2021,LXLWBLYHL2022,LXBLLLH2022}. For example, the Square Kilometre Array (SKA)~\cite{BBGKW2015,LXLWBLYHL2022} and other next-generation radio observatories (see~\cite{Murphy2018,RD2020}, and references therein) aim at an angular accuracy of about $1\mu$as. With the steady progress made in high-accuracy position, time, and angular measurements and the increase in the muiti-messenger synergic observations (see, e.g.,~\cite{BT2017,MFHM2019,IceCube2018,QJFZZ2021,HXJL2024}), a further consideration of gravitational lensing due to a moving lens becomes more and more meaningful and thus deserves our efforts.

However, to the best of our knowledge, most of the related works were devoted to the discussion of the velocity effects on the propagation of test particles constrained in the equatorial plane of the lens, and the full influence of the gravitational source's motion on the lensing properties of lightlike or timelike signals travelling off the lens' equatorial plane~\cite{Sereno2006,GY2010,AKP2011,JXJ2023} has not been considered to date. As a first step of a full theoretical treatment of it, here we mainly focus on probing the velocity effects on the gravitational deflection of light in the polar-axis plane of a moving KN lens. Additionally, the handling of this issue requires fundamentally the knowledge of the gravitational field of the background spacetime. Hence, the achievement of the metric of a moving KN black hole becomes another part of our concern naturally.

In present paper, we utilize the Lorentz boosting technique~\cite{PB1993,Klioner2003,WS2004} to derive the Kerr-Schild form of the exact metric of a constantly moving KN black hole, which is regarded as an extension
of a series of exact metrics of moving black holes~\cite{HL2014a,HL2014b,LFZMH2021}. Our attention is then concentrated on the investigation of the gravitational deflection effect of light restricted in the polar-axis plane of a moving KN black hole with its constant velocity along the polar axis within the 2PM approximation. It is interesting to find that this deflection angle is spin-independent. Moreover, the effect of the lens' motion on
the gravitational deflection of light is discussed and the possibility to detect the resulted velocity effect is evaluated. Our discussions are limited in the weak-field, small-angle, and thin-lens approximation.

The outline of this paper is as follows. We derive the Kerr-Schild form of the exact metric of a constantly moving KN black hole in Sec.~\ref{sect2}. As an application of the weak field limit of this solution, the null gravitational deflection in the polar-axis plane of a moving KN black hole with a constant velocity along the polar axis is probed in the 2PM approximation in Sec.~\ref{sect3}, in accompany with a discussion of the velocity effect on the deflection angle. Section~\ref{sect5} presents a summary subsequently. Throughout this paper, the metric signature $(-,~+,~+,~+)$ and the geometrized units in which $G = c = 1$ are adopted, Greek indices run over $0,~1,~2$, and $3$, and Latin indices run over $1,~2$, and $3$.

\section{The metric of a constantly moving KN black hole} \label{sect2}
Let $\bm{e}_i \left(i=1,~2,~3\right)$ denote the orthonormal basis of a three-dimensional Cartesian coordinate system. $x^\nu=\left(t,~x_1,~x_2,~x_3\right)$ and $X^\nu\equiv(T,~X_1,~X_2,~X_3)=(T,~\bm{X})$ stand for the rest Kerr-Schild coordinate frame of the observer of the background and the comoving Kerr-Schild coordinate frame of the gravitational source, respectively. The gravitational field outside a constantly rotating and charged black hole is described by the KN metric, which reads in the comoving Kerr-Schild frame~\cite{DKS1969,ACV1996,Xulu2000,NA2014}:
\begin{widetext}
\begin{eqnarray}
&&\nn \mathrm{d}s^2=-\,\mathrm{d}T^2+\mathrm{d}X_1^2+\mathrm{d}X_2^2+\mathrm{d}X_3^2+\frac{(2MR-Q^2)R^2}{R^4+a^2Z^2} \\
&&\hspace*{28pt}\times\left[\mathrm{d}T+\frac{X_3\mathrm{d}X_3}{R}+\frac{R\left(X_1\mathrm{d}X_1+X_2\mathrm{d}X_2\right)-a\left(X_1\mathrm{d}X_2-X_2\mathrm{d}X_1\right)}{R^2+a^2}\right]^2~, ~~~~~~ \label{metric-1}
\end{eqnarray}
\end{widetext}
where $R$ is defined by the relation $\frac{X_1^2+X_2^2}{R^2+a^2}+\frac{X_3^2}{R^2}=1$, and $M$, $Q$, and $a \left(\equiv |\bm{J}|/M\right)$ represent the rest mass, electric charge, and angular momentum per unit mass of the KN black hole, respectively, with $\bm{J}\equiv J\bm{e}_3$ along the polar axis.

Based on the Lorentz boosting technique~\cite{WS2004} and Eq.~\eqref{metric-1}, the external gravitational field of a moving KN black hole can be obtained. We know that the Lorentz transformation between the observer's rest frame, in which a KN black hole moves with an arbitrary constant velocity $\bm{v}=v_1\bm{e}_1+v_2\bm{e}_2+v_3\bm{e}_3$, and the comoving frame of the black hole has the form
\begin{equation}
X^\rho= \Lambda^\rho_\sigma x^\sigma~, \label{LT0}
\end{equation}
where
\begin{eqnarray}
&&\Lambda^0_0=\gamma~, \label{Lambda1}  \\
&&\Lambda^i_0=\Lambda^0_i=-v_i\gamma~, \label{Lambda2}  \\
&&\Lambda^i_j=\delta_{ij}+\frac{(\gamma-1)v_iv_j}{v^2}~.   \label{Lambda3}
\end{eqnarray}
Here, $v\equiv|\bm{v}|=\sqrt{v_1^2+v_2^2+v_3^2}$, and $\gamma\equiv(1-v^2)^{-\scriptstyle \frac{1}{2}}$ denotes the Lorentz factor. By applying the definition of the covariant metric tensor, the exact metric of the moving KN black hole can be achieved in the observer's rest Kerr-Schild frame $\left(t,~x_1,~x_2,~x_3\right)$ as follows:
\begin{widetext}
\begin{eqnarray}
&&g_{00}=-1+\frac{\gamma^2(2MR-Q^2)R^2}{R^4+a^2X_3^2}\!\left[1-\frac{v_1(RX_1+aX_2)}{R^2+a^2}-\frac{v_2(RX_2-aX_1)}{R^2+a^2}-\frac{v_3X_3}{R}\right]^2~,       \label{g00mKN}     \\
\nn&&g_{0i}=\frac{\gamma\left(2MR-Q^2\right)R^2}{R^4+a^2X_3^2}\!\left[1-\frac{v_1(RX_1\!+\!aX_2)}{R^2+a^2}-\frac{v_2(RX_2\!-\!aX_1)}{R^2+a^2}
-\frac{v_3X_3}{R}\right]\!\Bigg{\{}\frac{\left(RX_1\!+\!aX_2\right)\delta_{1i}}{R^2+a^2}    \\
&&\hspace*{25pt}+\,\frac{\!\left(RX_2\!-\!aX_1\right)\!\delta_{2i}}{R^2+a^2}\!+\!\frac{X_3\delta_{3i}}{R}\!-\!v_i\gamma
\!+\!\frac{v_i\!\left(\gamma\!-\!1\right)\!}{v^2}\!\left[\frac{v_1\!\left(RX_1\!+\!aX_2\right)}{R^2+a^2}
\!+\!\frac{v_2\!\left(RX_2\!-\!aX_1\right)\!}{R^2+a^2}\!+\!\frac{v_3X_3}{R}\right]\!\!\Bigg{\}}~,~~~~\label{g0imKN}
\end{eqnarray}
\begin{eqnarray}
\nn&&g_{ij}=\delta_{ij}+\frac{\left(2MR\!-\!Q^2\right)R^2}{R^4+a^2X_3^2}\bigg{\{}\frac{(RX_1\!+\!aX_2)\delta_{1i}}{R^2+a^2}
+\frac{\left(RX_2\!-\!aX_1\right)\delta_{2i}}{R^2+a^2}+\frac{X_3\delta_{3i}}{R}-v_i\gamma+\frac{v_i\left(\gamma\!-\!1\right)}{v^2} ~~~~~~~    \\
\nn&&\hspace*{24pt}\times\!\left[\frac{v_1\left(RX_1\!+\!aX_2\right)}{R^2+a^2}+\frac{v_2\left(RX_2\!-\!aX_1\right)}{R^2+a^2}+\frac{v_3X_3}{R}\right]\!\!\bigg{\}}
\!\left\{\frac{\left(RX_1\!+\!aX_2\right)\delta_{1j}}{R^2+a^2}+\frac{\!\left(RX_2\!-\!aX_1\right)\delta_{2j}}{R^2+a^2} \right. \\
&&\left.\hspace*{23pt}+\,\frac{X_3\delta_{3j}}{R}-v_j\gamma+\frac{v_j\left(\gamma-1\right)}{v^2}\!\left[\frac{v_1\left(RX_1+aX_2\right)}{R^2+a^2}
+\frac{v_2\left(RX_2-aX_1\right)}{R^2+a^2}+\frac{v_3X_3}{R}\right]\right\}~,~~  \label{gijmKN}
\end{eqnarray}
where $\delta_{ij}$ denotes the Kronecker delta. We can see that Eqs.~\eqref{g00mKN} - \eqref{gijmKN} for the case of no rotation of the black hole reduce to the Kerr-Schild form of the exact metric of a constantly moving Reissner-Nordstr\"{o}m source
\begin{eqnarray}
&&g_{00}=-1+\frac{\gamma^2(2MR-Q^2)}{R^2}\!\left(1-\frac{\bm{v}\cdot\bm{X}}{R}\right)^2~,       \label{g00mRN}     \\
&&g_{0i}=\frac{\gamma (2MR-Q^2)}{R^2}\!\left(1-\frac{\bm{v}\cdot\bm{X}}{R}\right)\!\left[\frac{X_i}{R}-v_i\gamma+\frac{v_i(\gamma-1)(\bm{v}\cdot\bm{X})}{v^2R}\right]~,~~~~\label{g0imRN}    \\
&&g_{ij}=\delta_{ij}+\frac{2MR-Q^2}{R^2}\!\left[\frac{X_i}{R}-v_i\gamma+\frac{v_i(\gamma-1)(\bm{v}\cdot\bm{X})}{v^2R}\right]
\!\left[\frac{X_j}{R}-v_j\gamma+\frac{v_j(\gamma-1)(\bm{v}\cdot\bm{X})}{v^2R}\right]~,~~~~\label{gijmRN}
\end{eqnarray}
where the distance between the gravitational source and the field point is reduced to $R=|\bm{X}|=\sqrt{X_1^2+X_2^2+X_3^2}$.

\section{Weak-field deflection in polar-axis plane of a KN black hole moving along the polar axis} \label{sect3}
In this section, we first derive the weak-field geodesic equations of test particles constrained in the polar-axis plane $\left(x_1=\partial/\partial x_1=0\right)$ of a moving KN black hole
whose velocity is assumed to be along the polar axis (i.e., $\bm{v}=v_3\bm{e}_3$) for simplicity, and then calculate the null gravitational deflection up to the 2PM order in the polar-axis plane of this moving lens. A brief discussion of the velocity effect on the deflection angle is given subsequently.

\subsection{The metric of the moving KN source up to the 2PM order}  \label{sect3A}
In order to calculate the polar-axis-plane equations of motion in the gravitational field of the moving lens up to the 2PM order, we need the weak-field form of the spacetime metric, which can be expanded from Eqs.~\eqref{g00mKN} - \eqref{gijmKN} in the from
\begin{eqnarray}
&&g_{00}=-1+\frac{\gamma_P^2\left(2MR-Q^2\right)}{R^2}\!\left(1-\frac{v_3X_3}{R}\right)^2+\mathcal{O}\left(M^3\right)~, \label{g00ms2-WF}  \\
&&g_{0i}=\gamma_P\left(1-\frac{v_3X_3}{R}\right)\left\{\frac{2MR-Q^2}{R^2}\left[\frac{X_i}{R}
+\left(\frac{\left(\gamma_P-1\right)X_3}{R}-v_3\gamma_P\right)\delta_{3i}\right]+\frac{2aMX_2\delta_{1i}}{R^3}\right\}+\mathcal{O}\left(M^3\right)~,~~~~~~~ \label{g0ims2-WF}  \\
\nn&&g_{ij}=\delta_{ij}+\frac{2MR-Q^2}{R^2}\!\left[\frac{X_i}{R}+\left(\frac{\left(\gamma_P-1\right)X_3}{R}-v_3\gamma_P\right)\delta_{3i}\right]
\!\left[\frac{X_j}{R}+\left(\frac{\left(\gamma_P-1\right)X_3}{R}-v_3\gamma_P\right)\delta_{3j}\right] \\
&&\hspace*{24pt}+\,\frac{2aMX_2}{R^3}\!\left\{\left[\frac{X_{j}}{R}+\left(\frac{\left(\gamma_P-1\right)\!X_3}{R}-v_3\gamma_P\right)\delta_{3j}\right]\!\delta_{1i}
+\left[\frac{X_{i}}{R}+\left(\frac{\left(\gamma_P-1\right)X_3}{R}-v_3\gamma_P\right)\delta_{3i}\right]\delta_{1j}\right\}+\mathcal{O}\left(M^3\right)~,~~~~~~ \label{gijmS-WF}
\end{eqnarray}
where $\gamma_P=(1-v_3^2)^{-\scriptstyle\frac{1}{2}}$ and $R=\sqrt{X_2^2+X_3^2}$, with a reduced Lorentz transformation given by
\begin{eqnarray}
X_2=x_2~, \hspace*{35pt} X_3=\gamma_P\left(x_3\!-\!v_3t\right)~, \hspace*{35pt} T=\gamma_P\left(t\!-\!v_3x_3\right)~.~~~~~~~   \label{LT2}
\end{eqnarray}

\subsection{Equations of motion in the polar-axis plane of the moving black hole}  \label{sect3B}
Based on Eqs.~\eqref{g00ms2-WF} - \eqref{gijmS-WF}, we can get the 2PM geodesic equations of light constrained in the polar-axis plane of the moving KN lens
\begin{eqnarray}
\nn&& 0=\ddot{t}\!-\!\gamma_P^3\bigg{\{}\frac{v_3M\left[\left(R\!-\!v_3X_3\right)X_3\!+\!2v_3x_2^2\right]}{R^5}
-\!\frac{2M^2\left(R\!-\!v_3X_3\right)^2}{R^6}\!-\!\frac{v_3Q^2\left[\left(R\!-\!v_3X_3\right)X_3\!+\!v_3x_2^2\right]}{R^6}\!\bigg{\}} \\
\nn&&\times\left(R-v_3X_3\right)\dot{t}^2\!+\!\gamma_P^3\bigg{\{}\!\frac{M\!\left[2\left(X_3^2\!-\!x_2^2\right)R\!-\!v_3\left(3X_3^2\!-\!v_3^2R^2\right)X_3\right]}{R^5}
\!+\!\frac{2M^2\left(X_3\!-\!v_3R\right)^2\left(R\!-\!v_3X_3\right)}{R^6} \\
\nn&&+\,\frac{Q^2\left[v_3\gamma_P^{-2}R^2X_3-2\left(R-v_3X_3\right)X_3^2+R\,x_2^2\right]}{R^6}\bigg{\}}\,\dot{x}_3^2+\frac{2\gamma_P^2M\left(R-3v_3X_3\right)\left(R-v_3X_3\right)x_2\,\dot{t}\,\dot{x}_2}{R^5} \\
\nn&&+\,2\gamma_P^3\bigg{\{}\hspace*{-1.2pt}\frac{M\!\left[\left(R\!-\!v_3X_3\right)\!X_3\!+\!2v_3x_2^2\right]}{R^5}\!+\!\frac{2M^2\!\left(X_3\!-\!v_3R\right)\left(R\!-\!v_3X_3\right)}{R^6}
\!-\!\frac{Q^2\!\left[\left(R\!-\!v_3X_3\right)\!X_3\!+\!v_3x_2^2\right]}{R^6}\bigg{\}} \\
&&\times\left(R-v_3X_3\right)\dot{t}\dot{x}_3+\frac{2\gamma_P^2M\left[4\left(R-v_3X_3\right)X_3-v_3x_2^2\right]x_2\dot{x}_2\dot{x}_3}{R^5}+\mathcal{O}\left(M^3\right)~,~~ \label{MP-t}
\end{eqnarray}
\begin{eqnarray}
&&\nn 0=\ddot{x}_2+\bigg{\{}\!\frac{M\!\left[X_3^2+\left(3\gamma_P^2-2\right)x_2^2\right]}{R^5}-\frac{2\gamma_P^2M^2\!\left(R-v_3X_3\right)^2}{R^6}
-\frac{Q^2\!\left[X_3^2+\left(2\gamma_P^2-1\right)x_2^2\right]}{R^6}\bigg{\}}x_2\dot{t}^2 \\
\nn&&-\!\left[\frac{M\!\left(R^2-3\gamma_P^2x_2^2\right)}{R^5}+\frac{2\gamma_P^2M^2\!\left(X_3-v_3R\right)^2}{R^6}
-\frac{Q^2\!\left(R^2-2\gamma_P^2x_2^2\right)}{R^6}\right]\!x_2\dot{x}_3^2\!+\!\frac{6v_3\gamma_PMX_3x_2^2~\dot{t}\,\dot{x}_2}{R^5} \\
\nn&&-\,2\,\gamma_P^2\left[\frac{3\,v_3Mx_2^2}{R^5}+\frac{2M^2\left(X_3-v_3R\right)\left(R-v_3X_3\right)}{R^6}-\frac{2\,v_3\,Q^2x_2^2}{R^6}\right]\!x_2\,\dot{t}\,\dot{x}_3
-\frac{6\gamma_P MX_3\,x_2^2\,\dot{x}_2\,\dot{x}_3}{R^5} \\
&&+\,\mathcal{O}\left(M^3\right)~, \label{MP-y}
\end{eqnarray}
\begin{eqnarray}
&&\nn 0=\ddot{x}_3+\gamma_P^3\bigg{\{}\frac{M\left[\,\left(R^2-3v_3^2X_3^2\right)X_3+2v_3^3\left(X_3^2-x_2^2\right)R\,\right]}{R^5}-\frac{2M^2\left(X_3-v_3R\right)\left(R-v_3X_3\right)^2}{R^6} \\
\nn&&-\frac{Q^2\!\left[\gamma_P^{-2}R^2X_3\!-\!2v_3^2\left(X_3\!-\!v_3R\right)\!X_3^2\!-\!v_3^3Rx_2^2\right]}{R^6}\!\bigg{\}}\dot{t}^2
\!-\!\gamma_P^3\left(X_3\!-\!v_3R\right)\!\bigg{\{}\!\frac{M\!\left[\left(X_3\!-\!v_3R\right)X_3\!-\!2x_2^2\right]}{R^5} \\
\nn&&+\frac{2M^2\left(X_3\!-\!v_3R\right)^2}{R^6}\!-\!\frac{Q^2\left[\left(X_3\!-\!v_3R\right)X_3\!-\!x_2^2\right]}{R^6}\bigg{\}}\dot{x}_3^2
+\frac{2\,v_3\gamma_P^2M\!\left[4\left(X_3\!-\!v_3R\right)X_3\!+\!x_2^2\right]x_2\dot{t}\dot{x}_2}{R^5} \\
\nn&&+\,2\gamma_P^3\left(X_3-v_3R\right)\bigg{\{}\frac{v_3M\left[\left(X_3-v_3R\right)X_3-2x_2^2\right]}{R^5}-\frac{2M^2\left(X_3-v_3R\right)\left(R-v_3X_3\right)}{R^6}-\frac{v_3\,Q^2}{R^6} \\
&&\times\!\left[\left(X_3-v_3R\right)X_3-x_2^2\right]\!\bigg{\}}\,\dot{t}\dot{x}_3
-\frac{2\gamma_P^2M\left(3X_3-v_3R\right)\left(X_3-v_3R\right)x_2\,\dot{x}_2\dot{x}_3}{R^5}+\mathcal{O}\left(M^3\right)~,  \label{MP-z}
\end{eqnarray}
where a dot denotes the derivative with respect to the affine parameter $\xi$ which traces the trajectory of a light ray~\cite{WS2004,We1972}, and we have neglected the third- and higher-order terms such as the ones containing the factor $M\dot{x}_2^2$ or $M^2\dot{x}_2$, since $\dot{x}_2$ is regarded to be of the order of $M/b_P$~\cite{HL2016b} in which $b_P$ denotes the impact parameter. Interestingly, it is found that all of the spin-dependent contributions have vanished in Eqs.~\eqref{MP-t} - \eqref{MP-z}. Moreover, it should be mentioned that due to the frame dragging effect caused by the lens' rotation, the trajectory of light propagating in the polar-axis plane of the moving black hole, indicated by Eqs.~\eqref{MP-t} - \eqref{MP-z}, is regarded as the projection of a three-dimensional path of light.

\subsection{The 2PM gravitational deflection of light by the moving lens}  \label{sect3C}
Now we consider the gravitational deflection up to the 2PM order of light constrained in the polar-axis plane of the moving KN black hole with a velocity $\bm{v}=v_3\bm{e}_3$ along the polar axis. Figure~\ref{Figure1} presents the corresponding geometrical diagram for light propagating from the source $A$ to the receiver $B$ in the polar-axis plane of the moving KN lens.

\begin{figure*}
\centering
\begin{minipage}[b]{\textwidth}
\includegraphics[width=16cm]{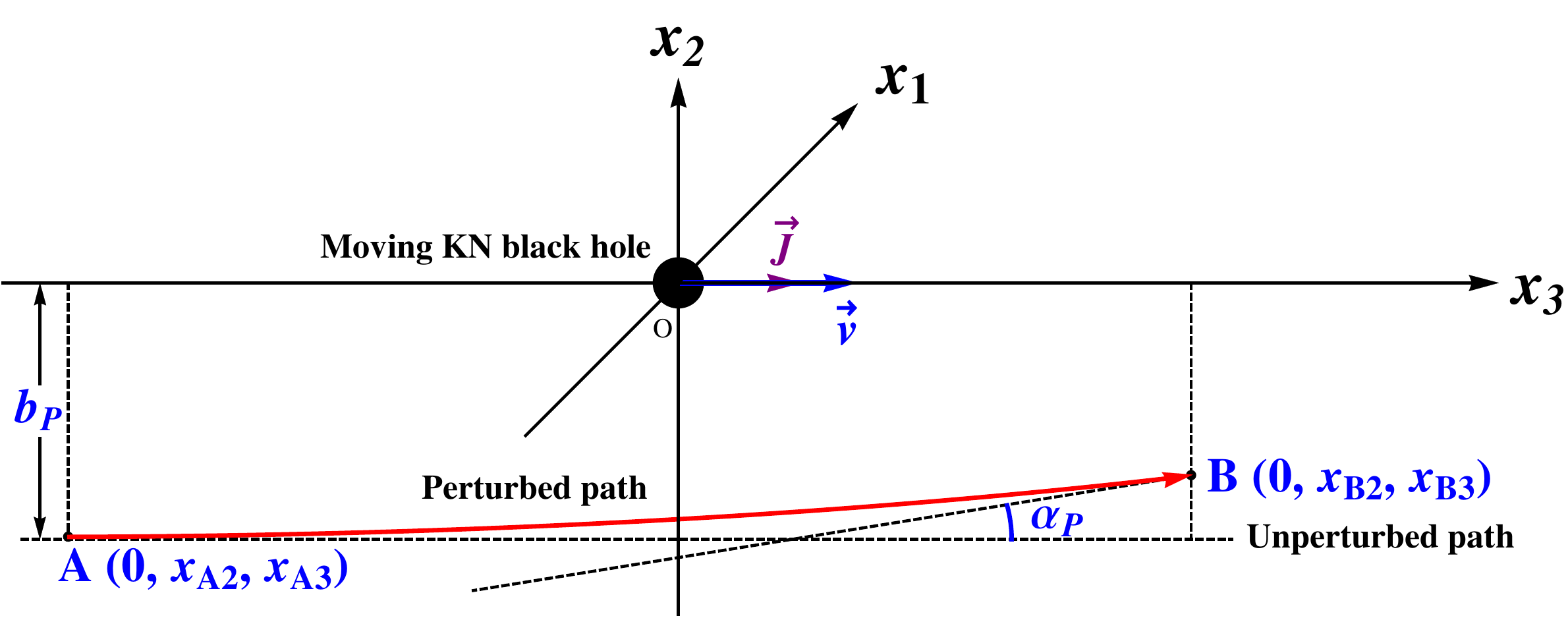}
\caption{Geometrical diagram for the propagation of light, which is emitted by the source $A$ and received by the detector $B$, constrained in the polar-axis plane of a moving KN lens whose angular momentum $\bm{J}$ and constant velocity $\bm{v}$ are along the $x_3$-axis. The red line stands for the propagation trajectory of light travelling from $x_3\rightarrow -\infty$ with an initial velocity $\bm{w}=\bm{e}_3$, $b_P$ is the impact parameter, and $\alpha_P$ denotes the deflection angle. The coordinates of the source and detector of light in the background's rest frame are denoted, respectively, by $(0,~x_{A2},~x_{A3})$ and $(0,~x_{B2},~x_{B3})$, with $x_{A2}\approx-b_P$, $x_{A3}\ll-b_P$, and $x_{B3}\gg b_P$. Their coordinates in the comoving frame are denoted by $(0,~X_{A2},~X_{A3})$ and $(0,~X_{B2},~X_{B3})$, respectively. The gravitational bending effect is exaggerated greatly to distinguish the perturbed propagating path (the red line) from the unperturbed one (the dashed horizontal line). }   \label{Figure1}
\end{minipage}
\end{figure*}

We then calculate the polar-axis-plane gravitational deflection of light up to the 2PM order iteratively via the approach developed in~\cite{WS2004,HL2017b}. A gravitational bending angle of a test particle is the difference of the propagation directions of the test particle, and is defined in our scenario by~\cite{HL2017b}
\begin{eqnarray}
\alpha_P\equiv\arctan\!\left.\frac{\mathrm{d}x_2}{\mathrm{d}x_3}\right|^{\xi\rightarrow +\infty}_{\xi\rightarrow -\infty}
=\arctan\!\left.\frac{\dot{x}_2}{\dot{x}_3}\right|^{x_3\rightarrow +\infty}_{x_3\rightarrow -\infty}
=\left.\frac{\dot{x_2}}{\dot{x_3}}\right|^B_A+\mathcal{O}\left(M^3\right)~,~~~     \label{angle-1}
\end{eqnarray}
\end{widetext}
where the third- and higher-order contributions to the deflection effect have been omitted in the last equality.

By using the boundary conditions $\dot{t}|_{\xi\rightarrow -\infty}=\dot{t}|_{x_3\rightarrow -\infty}=1$, $\dot{x}_2|_{\xi\rightarrow -\infty}=\dot{x}_2|_{x_3\rightarrow -\infty}=0$, and $\dot{x_3}|_{\xi\rightarrow -\infty}=\dot{x_3}|_{x_3\rightarrow -\infty}=1$ and assuming that $\xi$ has the dimension of length~\cite{WS2004}, we get the following zeroth-order expressions from Eqs.~\eqref{MP-t} - \eqref{MP-z}:
\begin{eqnarray}
&& \dot{t}=1+\mathcal{O}\left(M\right)~,      \label{Z0PM-dott}  \\
&& \dot{x}_2=0+\mathcal{O}\left(M\right)~,    \label{Z0PM-doty}  \\
&& \dot{x}_3=1+\mathcal{O}\left(M\right)~.    \label{Z0PM-dotz}
\end{eqnarray}
With the help of Eq.~\eqref{LT2} and the boundary condition $x_2|_{\xi\rightarrow -\infty}=x_2|_{x_3\rightarrow -\infty}=-b_P$, Eqs.~\eqref{Z0PM-dott} - \eqref{Z0PM-dotz} further imply the analytical forms of the coordinate $x_2$ and the parameter and coordinate transformations up to the 0PM order
\begin{eqnarray}
&&x_2=-b_P+\mathcal{O}\left(M\right)~,                                 \label{Z0PM-y}   \\
&&\mathrm{d}\xi=\left[1+\mathcal{O}\left(M\right)\right]\mathrm{d}x_3~,                \label{Z0PM-PT}  \\
&&\mathrm{d}x_3=\left[(1-v_3)^{-1}\gamma_P^{-1}+\mathcal{O}\left(M\right)\right]\mathrm{d}X_3~.      \label{Z0PM-CT}
\end{eqnarray}

Now we substitute Eqs.~\eqref{Z0PM-dott} - \eqref{Z0PM-CT} into Eqs.~\eqref{MP-t} - \eqref{MP-z} and thus obtain
\begin{widetext}
\begin{eqnarray}
&&\dot{t}=1+\frac{M\!\left[2(X_3^2+b_P^2)+2X_3\sqrt{X_3^2+b_P^2}+v_3b_P^2\right]}{(1+v_3)(X_3^2+b_P^2)^{\frac{3}{2}}}+\mathcal{O}\left(M^2\right)~,          \label{Z1PM-dott}  \\
&&\dot{x}_2=\frac{\left(1-v_3\right)\gamma_PMX_3\left(2X_3^2+3b_P^2\right)}{b_P\left(X_3^2+b_P^2\right)^{\frac{3}{2}}}+\mathcal{O}
\left(M^2\right)~,                  \label{Z1PM-doty}  \\
&&\dot{x}_3=1+\frac{M\left[b_P^2+2v_3\sqrt{X_3^2+b_P^2}\left(\sqrt{X_3^2+b_P^2}+X_3\right)\right]}{\left(1+v_3\right)\!\left(X_3^2+b_P^2\right)^{\frac{3}{2}}}
+\mathcal{O}\left(M^2\right)~.     \label{Z1PM-dotz}
\end{eqnarray}
Additionally, integrating Eq.~\eqref{Z1PM-doty} over $\xi$ and using Eqs.~\eqref{Z0PM-y} - \eqref{Z0PM-CT}, we can get the 1PM expression of $x_2$
\begin{eqnarray}
x_2=-b_P\!\left[1-\frac{M\left(2X_3^2+b_P^2\right)}{b_P^2\sqrt{X_3^2+b_P^2}}\right]+\mathcal{O}\left(M^2\right)~.            \label{Z1PM-y}
\end{eqnarray}
The analytical form of the parameter transformation up to the 1PM order can be obtained directly from Eq.~\eqref{Z1PM-dotz} as
\begin{eqnarray}
\mathrm{d}\xi=\left\{1-\frac{M\left[b_P^2+2v_3\sqrt{X_3^2+b_P^2}\left(\sqrt{X_3^2+b_P^2}+X_3\right)\right]}{\left(1+v_3\right)\left(X_3^2+b_P^2\right)^\frac{3}{2}}
+\mathcal{O}\left(M^2\right)\right\}\mathrm{d}x_3~.~~~~    \label{Z1PM-PT}
\end{eqnarray}
Simultaneously, the association of Eqs.~\eqref{Z1PM-dott} and~\eqref{Z1PM-dotz} with the reduced Lorentz transfromation shown in Eq.~\eqref{LT2} implies the 1PM expression of the coordinate transformation:
\begin{eqnarray}
\mathrm{d}x_3=\frac{1}{\left(1-v_3\right)\gamma_P}\!\left[1+\frac{v_3M\left(\sqrt{X_3^2+b_P^2}+X_3\right)^2}{\left(1+v_3\right)
\left(X_3^2+b_P^2\right)^{\frac{3}{2}}}+\mathcal{O}\left(M^2\right)\right]\mathrm{d}X_3~.~~~~           \label{Z1PM-CT}
\end{eqnarray}
\end{widetext}

We next substitute Eqs.~\eqref{Z1PM-dott} - \eqref{Z1PM-y} into the integration of Eq.~\eqref{MP-y} over $\xi$, adopt the 1PM parameter and coordinate transformations given in Eqs.~\eqref{Z1PM-PT} - \eqref{Z1PM-CT}, and obtain up to the 2PM order
\begin{widetext}
\begin{eqnarray}
&&\nn \dot{x}_2=\left(1-v_3\right)\gamma_P\Bigg{\{}\frac{M\left(2X_3^2+3b_P^2\right)X_3}{\left(X_3^2+b_P^2\right)^\frac{3}{2}b_P}
-\frac{Q^2\!\left[5X_3b_P^3+3X_3^3b_P+3\left(X_3^2+b_P^2\right)^2\arctan\frac{X_3}{b_P}\right]}{4\left(X_3^2+b_P^2\right)^2b_P^2}    \\
&&\hspace*{22.5pt}+\,\frac{3M^2\!\left[5X_3^5b_P+16X_3^3b_P^3+7X_3b_P^5+5\left(X_3^2+b_P^2\right)^3\arctan\frac{X_3}{b_P}\right]}{4\left(X_3^2+b_P^2\right)^3b_P^2}\Bigg{\}}+\mathcal{O}\left(M^3\right)~. \label{Z2PM-doty}
\end{eqnarray}
\end{widetext}
Eventually, the substitution of Eqs.~\eqref{Z1PM-dotz} and \eqref{Z2PM-doty} into Eq.~\eqref{angle-1}, along with the conditions $X_{A3}\ll -b_P~\left(\approx X_{A2}\right)$ and $X_{B3}\gg b_P$ in the comoving Kerr-Schild frame, gives the gravitational deflection angle up to the 2PM order for light propagating from the source $A$ to the receiver $B$ in the polar-axis plane of the moving KN lens, which is expressed in the observer's rest
Kerr-Schild frame as
\begin{widetext}
\begin{equation}
\alpha_P=\left(1-v_3\right)\gamma_P\left(\frac{4M}{b_P}+\frac{15\pi M^2}{4b_P^2}-\frac{3\pi Q^2}{4b_P^2}\right)+\mathcal{O}\left(M^3\right)~.    \label{PDA-2PM}
\end{equation}
\end{widetext}

\begin{widetext}
\begin{figure*}
\centering
\begin{minipage}[b]{\textwidth}
\includegraphics[width=15cm]{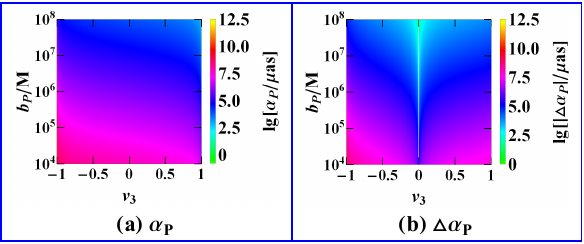}
\caption{The polar-axis-plane deflection angle $\alpha_P$ and the velocity effect $\Delta\alpha_P$ plotted for various $v_3$ and $b_P$ in color-indexed form.
Here a weak electrical charge $Q=0.01M$ of the moving lens is assumed. }   \label{Figure2}
\end{minipage}
\end{figure*}
\end{widetext}

\begin{table}
\begin{minipage}[t]{0.5\textwidth}
  \centering
\begin{tabular}{cccccccc} \toprule[1.2px]
    $v_3\:\backslash\:\text{Observables}$    &  ~~~~~~~~~~~~$\alpha_P$ ~~~~~~~~~~~~ &   ~~~~~~~~~$\Delta\alpha_P$~~~~~~~~~~            \\   \midrule[0.5pt]
                               0.99         &        5.85$\times$10$^5$        &           $-7.67\times$10$^6$                    \\
                                0.9         &        1.89$\times$10$^6$        &           $-6.36\times$10$^6$                    \\
                                0.1         &        7.46$\times$10$^6$        &           $-7.88\times$10$^5$                    \\
                              0.001         &        8.24$\times$10$^6$        &           $-8.25\times$10$^3$                    \\
                            0.00001         &        8.25$\times$10$^6$        &                $-82.51$                          \\
                          0.0000001         &        8.25$\times$10$^6$        &                 $-0.83$                          \\
                       $-$0.0000001         &        8.25$\times$10$^6$        &                  $0.83$                          \\
                         $-$0.00001         &        8.25$\times$10$^6$        &                 $82.51$                          \\
                           $-$0.001         &        8.26$\times$10$^6$        &            8.25$\times$10$^3$                    \\
                             $-$0.1         &        9.12$\times$10$^6$        &            8.71$\times$10$^5$                    \\
                             $-$0.9         &        3.60$\times$10$^7$        &            2.77$\times$10$^7$                    \\
                            $-$0.99         &        1.16$\times$10$^8$        &            1.08$\times$10$^8$                    \\   \bottomrule[1.2px]
\end{tabular} \par  \vspace*{3pt}
\end{minipage}
\caption{The values (in units of $\mu$as) of $\alpha_P$ and $\Delta\alpha_P$ for different polar-axis velocity $v_3$. A typical weak field $M/b_P=1.0\times10^{-5}$ and $Q=0.01M$ are assumed.  }   \label{Table1}
\end{table}

With respect to Eq.~\eqref{PDA-2PM}, three aspects should be mentioned. A first one is that the polar-axis-plane bending angle shown in Eq.~\eqref{PDA-2PM} is independent on the rotation or angular momentum of the moving KN black hole. Secondly, we find that for the case of no translational motion of the lens ($v_3=0$), Eq.~\eqref{PDA-2PM} matches well with the result of the null gravitational deflection in the polar-axis plane of a stationary KN lens derived in harmonic coordinates (see appendix~\ref{appendix-A} for details). Finally, the deflection angle $\alpha_P$ is plotted as the function of the impact parameter $b_P$ and the polar-axis velocity $v_3$ of the moving KN black hole in color-indexed form on the left of Fig.~\ref{Figure2}, and its values for a typical weak field $M/b_P=1.0\times10^{-5}$ are shown in Tab.~\ref{Table1}.

\subsection{Discussion of the velocity effect}  \label{sect3D}
As an example of potential astrophysical applications of the result, we assume that the moving KN black hole is resulted from the merger of two unequal-mass black holes, receives a recoil (or a ``kick") because of the asymmetric loss of linear momentum in the gravitational radiation~\cite{GSBHH2007}, and thus obtain a constant kick velocity $\bm{v}_3~(=v_3\bm{e}_3)$ along the polar axis. For discussing the influence of the lens' motion on the polar-axis-plane deflection more conveniently, we define a pure velocity-induced effect $\Delta\alpha_P\equiv\alpha_P-\alpha_P\!\left.\right|_{v_3=0}$, which means the deviation of $\alpha_P$ from the polar-axis-plane deflection angle due to a stationary KN source and is shown as the color-indexed function of $b_P$ and the polar-axis kick velocity $v_3$ in Fig.~\ref{Figure2}.

It indicates from Fig.~\ref{Figure2}, as well as Tab.~\ref{Table1} which contains some values of $\Delta\alpha_P$, that the velocity effect on the polar-axis-plane deflection angle increases monotonously with the decrease of the kick velocity from a positive ultra-relativistic value (i.e., $v_3\rightarrow1$) to a negative ultra-relativistic one (i.e., $v_3\rightarrow-1$). With respect to the detectability of $\Delta\alpha_P$, we find that its absolute values are so large that it is still possible to measure this velocity effect in near future (or even current) resolution when the lens takes a nonrelativistic motion at a low kick velocity.
For example, if the galactic supermassive black hole (i.e., Sgr A$^*$) with a rest mass $M=2.0\times10^{-10}\,$kpc~\cite{BG2016,Parsa2017} is assumed to be the lens and moves at a small kick velocity $v_3=0.0001$~($\sim\,$30km/s), then the velocity effect $\left|\Delta\alpha_P\right|$ is about $288.9\mu$as when we suppose the impact parameter $b_P$ to take a special value $5.74\times10^{-5}\,$kpc~\cite{HZFMWPL2020} of the Einstein radius of Sgr A$^*$. It is noticed that this value of $\Delta\alpha_P$ exceeds evidently the current multiwavelength astrometric precision (at the tens of $\mu$as level or better~\cite{Brown2021,Abuter2017a}), and
is much larger than the intended angular accuracy of next-generation radio observatories (such as the SKA~\cite{BBGKW2015,LXLWBLYHL2022}).

\section{Summary} \label{sect5}
In this work, we have focused on the investigation of the gravitational deflection effect up to the 2PM order for light constrained in the polar-axis plane of a moving Kerr-Newman black hole with a constant velocity along the polar axis. By means of the Lorentz boosting technique in Kerr-Schild coordinates, we have obtained the exact metric of a moving Kerr-Newman black hole with an arbitrary constant velocity for the first time.
The weak-field form of the resulting metric is then applied to the calculations of the dynamics of light, on the basis of which the null polar-axis-plane deflection angle induced by the moving Kerr-Newman lens
whose velocity is along the polar axis has been derived iteratively. It is interesting to find that the polar-axis-plane bending angle is independent on the intrinsic angular momentum of the rotating lens. The velocity-induced effect due to the lens' translational motion on the gravitational deflection has also been discussed. With the consideration that the study of propagation of light signals in time-dependent gravitational fields is important for modern relativistic astrophysics and fundamental astrometry, it should be expected that the results presented in this work might be helpful for future astronomical observations.

\section*{Acknowledgments}
This work was supported in part by the National Natural Science Foundation of China (Grant Nos. 11973025, 12205139, 12303079, 12475057, and 12481540180) and the Natural Science Foundation of Hunan Province (Grant No. 2022JJ40347). G.H. would like to thank Yi Xie for an enlightening discussion.

\appendix
\begin{widetext}
\section{Null gravitational deflection restricted in polar-axis plane of a stationary KN black hole in harmonic coordinates} \label{appendix-A}
The weak-field metric of a stationary KN black hole can be written in harmonic coordinates $x_H^\mu\equiv(t,~x_H,~y_H,~z_H)=(t,~\bm{x}_H)$ as follows:
\begin{eqnarray}
&&g_{00}=-1+\frac{2M}{r_H}-\frac{2M^2}{r_H^2}-\frac{Q^2}{r_H^2}+\mathcal{O}\left(M^3\right)~, \label{g00H}    \\
&&g_{0i}=\zeta_i+\mathcal{O}\left(M^3\right)~, \label{g0iH}   \\
&&g_{ij}=\left(1+\frac{M}{r_H}\right)^2\delta_{ij}+\frac{M^2-Q^2}{r_H^2}\frac{x_H^i x_{H}^j}{r_H^2}+\mathcal{O}\left(M^3\right)~, \label{gijH}
\end{eqnarray}
where $M$, $Q$, and $\delta_{ij}$ have been defined above, $r_H$ is related to $x_H$, $y_H$, and $z_H$ by $\frac{x_H^2+y_H^2}{r_H^2+a^2}+\frac{z_H^2}{r_H^2}=1$, and $\bm{\zeta}\equiv\frac{2aM}{r_H^3}\left(\bm{x}_H\times\bm{e_3}\right)=\left(\zeta_1,\zeta_2,0\right)$. Combining Eqs.~\eqref{g00H} - \eqref{gijH} and the geodesic equations $\frac{\mathrm{d}^{2} x^{\mu}}{\mathrm{d} \xi^{2}}+\Gamma^\mu_{\nu\lambda}\!\frac{\mathrm{d} x^{\nu}}{\mathrm{d} \xi}\!\frac{\mathrm{d} x^{\lambda}}{\mathrm{d} \xi}=0$ where $\Gamma^\mu_{\nu\lambda}$ is the Christoffel symbol,
we get the 2PM equations of motion for light constrained in the polar-axis plane $(x_H=\partial/\partial x_H=0)$ of the KN source
\begin{eqnarray}
&&0=\ddot{t}+\frac{2My_H\dot{t}\dot{y}_H}{r_H^3}+\frac{2\left(Mr_H-Q^2\right)z_H\dot{t}\dot{z}_H}{r_H^4}+\mathcal{O}\left(M^3\right)~, \label{GE-t}    \\
&&0=\ddot{y}_H+\frac{\left(Mr_H-4M^2-Q^2\right)y_H\dot{t}^2}{r_H^4}+\frac{\left[Mr_H^3-2M^2z_H^2+Q^2\left(z_H^2-y_H^2\right)\right]y_H\dot{z}_H^2}{r_H^6}
-\frac{2Mz_H\dot{y}_H\dot{z}_H}{r_H^3}+\mathcal{O}\left(M^3\right)~,~~~~~~~\label{GE-y}  \\
&&0=\ddot{z}_H+\frac{\left(Mr_H-4M^2-Q^2\right)z_H\dot{t}^2}{r_H^4}-\frac{\left[Mr_H^3-2M^2y_H^2+Q^2\left(y_H^2-z_H^2\right)\right]z_H\dot{z}_H^2}{r_H^6}
-\frac{2My_H\dot{y}_H\dot{z}_H}{r_H^3}+\mathcal{O}\left(M^3\right)~.~~~~~~~\label{GE-z}
\end{eqnarray}
\end{widetext}
Here, $r_H=\sqrt{y_H^2+z_H^2}$, a dot denotes the derivative with respect to $\xi$ above, and $\dot{y}_H$ has been regarded to be of the order of $M/b_P$~\cite{HL2016b}.

Similarly, we apply the iterative technique to the derivation of the gravitational deflection angle up to the 2PM order for light travelling in the polar-axis plane of the KN lens from the emitter $C~(0,~y_{CH},~z_{CH})$ to the receiver $D~(0,~y_{DH},~z_{DH})$, with $y_{CH}\approx-b_P$, $z_{CH}\ll-b_P$, and $z_{DH}\gg b_P$. By adopting the following boundary conditions
\begin{eqnarray}
&&\dot{t}|_{\xi\rightarrow -\infty}=\dot{t}|_{z_H\rightarrow -\infty}=1~,  \label{BC1-H}  \\
&&\dot{y}_H|_{\xi\rightarrow -\infty}=\dot{y}_H|_{z_H\rightarrow -\infty}=0~,  \label{BC2-H}  \\
&&\dot{z}_H|_{\xi\rightarrow -\infty}=\dot{z}_H|_{z_H\rightarrow -\infty}=1~,  \label{BC3-H}  \\
&&y_H|_{\xi\rightarrow -\infty}=y_H|_{z_H\rightarrow -\infty}=-b_P~,  \label{BC4-H}
\end{eqnarray}
we can obtain the analytical expressions up to the 1PM order for $\dot{t}$, $\dot{y}_H$, $\dot{z}_H$, $y_H$, and the parameter transformation from Eqs.~\eqref{GE-t} - \eqref{GE-z}
\begin{eqnarray}
&&\dot{t}=1+\frac{2M}{\sqrt{z_H^2+b_P^2}}+\mathcal{O}\left(M^2\right)~,                                            \label{z1PM-dott}  \\
&&\dot{y}_H=\frac{2Mz_H}{b_P\sqrt{z_H^2+b_P^2}}+\mathcal{O}\left(M^2\right)~,                                      \label{z1PM-doty}  \\
&&\dot{z}_H=1+\mathcal{O}\left(M^2\right)~,                                                                        \label{z1PM-dotz}  \\
&&y_H=-b_P\!\left(1-\frac{2M\sqrt{z_H^2+b_P^2}}{b_P^2}\right)+\mathcal{O}\left(M^2\right)~,~~~~~~                  \label{z1PM-y}     \\
&&\mathrm{d}\xi=\left[1+\mathcal{O}\left(M^2\right)\right]\mathrm{d}z_H~.                                                            \label{z1PM-PT}
\end{eqnarray}
Moreover, the substitution of Eqs.~\eqref{z1PM-dott} - \eqref{z1PM-PT} into the integration of Eq.~\eqref{GE-y} over $\xi$ implies
\begin{widetext}
\begin{eqnarray}
&&\nn\dot{y}_H\!=\!\frac{2Mz_H}{b_P\sqrt{z_H^2+b_P^2}}+\frac{M^2}{4b_P^2\left(z_H^2+b_P^2\right)^2}\!\left[17b_P^3z_H+15b_Pz_H^3+15\left(z_H^2+b_P^2\right)^2\arctan\!\frac{z_H}{b_P}\right] \\
&&\hspace*{23.5pt}-\,\frac{Q^2}{4b_P^2\left(z_H^2+b_P^2\right)^2}\!\left[5b_P^3z_H+3b_{P}z_H^3+3\left(z_H^2+b_P^2\right)^2\arctan\!\frac{z_H}{b_P}\right]+\mathcal{O}\left(M^3\right)~.   \label{z2PM-doty}
\end{eqnarray}

Subsequently, with the consideration of Eqs.~\eqref{z1PM-dotz} and \eqref{z2PM-doty} and the conditions $z_{CH}\ll-b_P\,\left(\approx y_{CH}\right)$ and $z_{DH}\gg b_P$, the deflection angle up to the 2PM order of light in the polar-axis plane a KN black hole takes the form in harmonic coordinates:
\begin{eqnarray}
\alpha_P=\arctan\!\left.\frac{\dot{y}_H}{\dot{z}_H}\right|^{z_H\rightarrow +\infty}_{z_H\rightarrow -\infty}
=\left.\frac{\dot{y}_H}{\dot{z}_H}\right|^{D}_{C}+\mathcal{O}\left(M^3\right)=\frac{4M}{b_P}+\frac{15\pi M^2}{4b_P^2}-\frac{3\pi Q^2}{4b_P^2}+\mathcal{O}\left(M^3\right)~,~~~~~~~    \label{HPDA-2PM}
\end{eqnarray}
which is the same as that given in \eqref{PDA-2PM} for the case of $v_3=0$.
\end{widetext}

\end{document}